# Bound states of a short-range potential with inverse cube singularity

A. D. Alhaidari

*Saudi Center for Theoretical Physics, P. O. Box 32741, Jeddah 21438, Saudi Arabia*

**Abstract**: We use the Tridiagonal Representation Approach (TRA) to obtain exact bound states solution (energy spectrum and wavefunction) of the Schrödinger equation for a three-parameter short-range potential with $1/r$, $1/r^2$ and $1/r^3$ singularities at the origin. The solution is a finite series of square integrable functions with weighted coefficients that satisfy a three-term recursion relation. The solution of the recursion is the discrete version of a non-conventional orthogonal polynomial. We are currently preparing to use the results of this work to study the binding of an electron to a molecule with an effective electric quadrupole moment, which has the same $1/r^3$ singularity.

**PACS**: 03.65.Ge, 03.65.Fd, 34.80.Bm, 03.65.Ca

**Keywords**: $1/r^3$ singular potential, tridiagonal representations, recursion relation, orthogonal polynomials

## 1. Introduction

Exact solution of the wave equation gives an analytic robust understanding of the corresponding physical system especially in situations where numerical solutions fail. Such cases may occur at critical couplings, strong singularities, limiting behavior or phase transitions. Therefore, finding exact solutions of the wave equation for as many potential models as possible continues to be one of the most challenging tasks in quantum mechanics. Many methods to achieve this goal has been devised over time leading to a well-known class of exactly solvable problems. Prominent examples of such systems include, but not limited to, the Coulomb, oscillator, Morse, Pöschl-Teller and Eckart potentials [1,2]. Attempts by many researchers to enlarge these classes of exactly solvable potentials are continuing. In 2005, we introduced an algebraic method for solving the wave equation based on the J-matrix method and the theory of orthogonal polynomials [3]. The method is now well established with rich mathematical underpinnings and referred to as the "Tridiagonal Representation Approach (TRA)". It turned out that the class of exactly solvable potentials using the TRA is larger than the well-known conventional class. We find either new solvable potentials or generalizations of known ones. As illustrative examples, we mention the infinite square well with sinusoidal bottom [4-6], the hyperbolic pulse (or hyperbolic single wave) [7,8] and a generalization of the Eckart potential by adding to it the short-range singular component $\frac{e^{-\lambda r}}{e^{\lambda r}-1}$ [9,10]. A partial list of this extended class of integrable potentials is shown as Table 1 in Ref. [11]. In this work, we use the TRA to obtain exact bound states solution (energy spectrum and wavefunction) of the S-wave Schrödinger equation for a three-parameter short-range potential with $1/r$, $1/r^2$ and $1/r^3$ singularities at the origin. It is another non-trivial and special generalization of the hyperbolic Ekart potential obtained by adding the $1/r^3$ singular term $\cosh(\lambda r)/\sinh^3(\lambda r)$. To the best of our knowledge, this would be the first published work on exact solution of such highly singular potential. A brief review of the TRA was published recently in Ref. [12]. The outline of the method goes as follows: The wavefunction is written as a series (finite or infinite) in terms of a complete set of square integrable basis functions. The basis is required to produce a tridiagonal matrix representation for the wave operator. Consequently, the matrix wave equation becomes a three-term recursion relation for the expansion coefficients of the series. We solve the recursion in terms of orthogonal polynomials whose analytic properties (e.g., weight function, asymptotics, generating function, spectrum formula, etc.) give the physical properties of the system such as the bound states energies, the density of states, the scattering phase shift, etc.

In the following section, we setup the problem following the TRA formulation. Then in section 3, we derive the bound states wavefunction as a finite series and identify the corresponding orthogonal polynomial. Unfortunately, this polynomial is found nowhere in the appropriate mathematics literature although it has already been encountered frequently in the physics literature. The polynomial could be written explicitly to any desired degree albeit not in closed form. We calculate the energy spectrum, which is finite, for a given physical configuration and plot the corresponding wavefunctions.

## 2. TRA formulation of the problem

In the atomic units $\hbar = m = 1$, the time-independent one-dimensional Schrödinger equation for a point particle of mass $m$ under the influence of a potential $V(r)$ reads as follows

$$\left[ -\frac{1}{2}\frac{d^2}{dr^2} + V(r) - E \right] |\psi(r)\rangle = 0, \tag{1}$$

where $r \geq 0$, $E$ is the particle energy and $\psi(r)$ its wavefunction. For bound states, the wavefunction vanishes at the origin and at infinity. Now, we make a transformation to a dimensionless coordinate $x(r) = \coth(\lambda r)$, where $\lambda$ is a real positive scale parameter. Thus, $x \geq 1$ and Eq. (1) is mapped into the following second order differential equation in terms of $x$

$$\lambda^2 (1-x^2) \left[ (1-x^2)\frac{d^2}{dx^2} - 2x\frac{d}{dx} + \frac{\varepsilon - U(x)}{1-x^2} \right] \psi(r(x)) = 0, \tag{2}$$

where $\varepsilon = 2E/\lambda^2$ and $U(x) = 2V(r)/\lambda^2$. In accordance with the TRA, we search for a complete set of square integrable basis $\{\phi_n(x)\}$ to expand the wavefunction in a series as $|\psi(r)\rangle = \sum_n f_n |\phi_n(x)\rangle$. Additionally, we require that the basis set carries a tridiagonal matrix representation for the wave operator $J = -\frac{1}{2}\frac{d^2}{dr^2} + V(r) - E$. Consequently, the matrix wave equation becomes a three-term recursion relation for the expansion coefficients $\{f_n\}$. The basis elements are to be written in terms of orthogonal polynomials whose argument is compatible with the range of the configuration space coordinate (i.e., $x \geq 1$). Now, the orthogonality properties of the Jacobi polynomials shown in Appendix A by Eq. (A6) suggests that we can choose the following set of functions as basis elements

$$\phi_n(x) = c_n (x-1)^\alpha (x+1)^{-\beta} P_n^{(\mu,\nu)}(x), \tag{3}$$

where $\mu > -1$, $\mu + \nu < -2N-1$, $n \in \{0,1,2,...,N\}$ and $N$ is some large enough non-negative integer. Therefore, $\nu$ must be negative whereas the normalization constant is suggested by the orthogonality relation (A6) as $c_n = \sqrt{(2n+\mu+\nu+1)\frac{\Gamma(n+1)\Gamma(n+\mu+\nu+1)}{\Gamma(n+\mu+1)\Gamma(n+\nu+1)}}$ $\sqrt{\frac{\sin\pi(\mu+\nu+1)}{2^{\mu+\nu+1}\sin\pi\nu}}$. The parameters $\alpha$ and $\beta$ will be determined below by the tridiagonal representation requirement. In terms of the variable $x$, the wave operator $J$ is written using Eq. (2) as follows

$$J = -\frac{\lambda^2}{2}(1-x^2)\left[(1-x^2)\frac{d^2}{dx^2} - 2x\frac{d}{dx} + \frac{\varepsilon - U(x)}{1-x^2}\right]. \tag{4}$$

Using the differential equation of the Jacobi polynomials (A2), we obtain the following action of the wave operator on the basis elements (3)

$$J\phi_n(x) = -\frac{\lambda^2}{2}c_n(1-x^2)(x-1)^\alpha (x+1)^{-\beta}\left\{\left[x(\mu+\nu-2\alpha+2\beta) - (2\alpha+2\beta+\nu-\mu)\right]\frac{d}{dx}\right.$$
$$\left. + \frac{2\alpha^2}{1-x} + \frac{2\beta^2}{1+x} - (\alpha-\beta)(\alpha-\beta+1) - n(n+\mu+\nu+1) + \frac{\varepsilon-U(x)}{1-x^2}\right\}P_n^{(\mu,\nu)}(x) \tag{5}$$

Using the differential property of the Jacobi polynomials (A3), this action becomes

$$J\phi_n(x) = -\frac{\lambda^2}{2}c_n(1-x^2)(x-1)^\alpha(x+1)^{-\beta} \times$$
$$\left\{\left(\frac{\mu-2\alpha}{1-x} - \frac{\nu+2\beta}{1+x}\right)\left[2\frac{(n+\mu)(n+\nu)}{2n+\mu+\nu}P_{n-1}^{(\mu,\nu)} - n\left(x+\frac{\nu-\mu}{2n+\mu+\nu}\right)P_n^{(\mu,\nu)}\right]\right. \tag{6}$$
$$\left. + \left[\frac{2\alpha^2}{1-x} + \frac{2\beta^2}{1+x} - (\alpha-\beta)(\alpha-\beta+1) - n(n+\mu+\nu+1) + \frac{\varepsilon-U(x)}{1-x^2}\right]P_n^{(\mu,\nu)}\right\}$$

To produce a tridiagonal representation for $J$, the right side of this equation must be a combination of $\phi_n(x)$ and $\phi_{n\pm 1}(x)$ with constant ($x$-independent) factors. The recursion relation of the Jacobi polynomials (A4) shows that (6) will contain terms proportional to $P_n^{(\mu,\nu)}$ and $P_{n\pm 1}^{(\mu,\nu)}$ with constant factors inside the curly brackets only in one of three cases:

(a) $2\alpha = \mu$, $2\beta = -\nu$, and $\dfrac{\varepsilon - U(x)}{1-x^2} = -\dfrac{2\alpha^2}{1-x} - \dfrac{2\beta^2}{1+x} - B + Cx$. \hfill (7a)

(b) $2\alpha = \mu$ and $\dfrac{\varepsilon - U(x)}{1-x} = -\dfrac{4\alpha^2}{1-x} - B + Cx$. \hfill (7b)

–3–

(c) $2\beta = -\nu$ and $\dfrac{\varepsilon - U(x)}{1+x} = -\dfrac{4\beta^2}{1+x} - B + Cx$. (7c)

where $B$ and $C$ are arbitrary real dimensionless constants. The terms with $\alpha^2$ and $\beta^2$ on the right side of the equations for $\varepsilon - U(x)$ are needed to cancel the corresponding terms in Eq. (6) that destroy the tridiagonal structure. The potential functions in the last two cases, (7b) and (7c), correspond to the hyperbolic Eckart potential, $V(r) = V_0 + V_1 \coth(\lambda r) + \dfrac{V_2}{\sinh^2(\lambda r)}$, which belongs to the well-known class of exactly solvable potentials [1,2]. Consequently, we ignore those two cases and focus on the first case where

$$\dfrac{U(x) - \varepsilon}{1-x^2} = \dfrac{\mu^2/2}{1-x} + \dfrac{\nu^2/2}{1+x} + B - Cx. \quad (8)$$

Therefore, we must choose $\mu^2 + \nu^2 = -2\varepsilon$ making the basis parameters energy dependent and limiting our exact solution to negative energies (i.e., bound states). Moreover, we obtain the following three-parameter potential function

$$\dfrac{2}{\lambda^2} V(r) = A \coth(\lambda r) - \dfrac{B}{\sinh^2(\lambda r)} + C \dfrac{\cosh(\lambda r)}{\sinh^3(\lambda r)}, \quad (9)$$

where $\mu^2 - \nu^2 = 2A$. The first two terms in this potential correspond to the hyperbolic Eckart potential. However, the last term is inverse cube singular at the origin and does not belong to the known class of exactly solvable potentials. Nonetheless, the TRA will produce an exact solution for this potential as we shall see below. Therefore, we devote our work here to the case where $C \neq 0$.

The potential (9) does not vanish at infinity but goes to the constant value $\tfrac{1}{2}\lambda^2 A$. This can be fixed by subtracting it out and writing

$$\dfrac{2}{\lambda^2} V(r) = A[\coth(\lambda r) - 1] - \dfrac{B}{\sinh^2(\lambda r)} + C \dfrac{\cosh(\lambda r)}{\sinh^3(\lambda r)}, \quad (10)$$

To accomplish that, we go back to Eq. (8) and make the replacement $U(x) \to U(x) + A$ and $\varepsilon \to \varepsilon + A$. As a result, we must choose $\mu^2 + \nu^2 = -2(\varepsilon + A)$. This modified relation together with $\mu^2 - \nu^2 = 2A$ give

$$\mu(\varepsilon) = \sqrt{-\varepsilon}, \; \nu(\varepsilon) = -\sqrt{-\varepsilon - 2A}, \quad (11)$$

requiring that either $\varepsilon < 0$ if $A$ is negative or $\varepsilon < -2A$ if $A$ is positive. However, the requirement that $\mu + \nu < -2N - 1$ for all $N$ selects the former condition dictating that $A$ must be negative. In fact, the existence of bound states (i.e., $N \geq 0$) gives a stronger constraint on the value of $A$, which is that $A \leq -\tfrac{1}{2}$. The number of basis elements (size of the basis), which is equal to $N+1$, depends on the potential parameter $A$ and is determined by the condition that $\mu+\nu<-2N-1$. It is not difficult to show that $N$ is the largest integer less than or equal to $\sqrt{-A/2}-\frac{1}{2}$. For a given negative value of $A$, Figure 1 shows the $\mu(\varepsilon)+\nu(\varepsilon)$ curve and several horizontal dashed lines corresponding to $-2k-1$ for $k=0,1,2,...$ The solid horizontal line is the $-2N-1$ line. Therefore, in our formulation of the problem with $C\neq 0$, bound state solution is possible only if $A\leq-\frac{1}{2}$. Moreover, the number of bound states (size of the energy spectrum) must, therefore, be finite and, in fact, it is less than or equal to the size of the basis $N$.

It should be obvious that the potential (10) is singular at the origin with $r^{-1}$, $r^{-2}$ and $r^{-3}$ singularities of strength $A$, $-B$ and $C$, respectively. However, it has a short range since it decays exponentially away from the origin as $e^{-2\lambda r}$. Thus, $\lambda$ gives a measure of the shortness of the range of this potential. It should also be clear that the shape of the potential is governed by two parameter ratios such as $A/C$ and $B/C$. Figure 2 shows a typical configuration of the potential for a given set of values of these two ratios. The Figure shows that it resembles the effective potential of the Coulomb problem with non-zero angular momentum and negative charge. However, there are three main difference. The first is that the strongest singularity of the potential (10) at the origin is $r^{-3}$ whereas it is $r^{-2}$ for the effective Coulomb potential. Second, the potential valley in the Coulomb problem is due to the kinetic energy (orbital term) for non-zero angular momentum whereas here it is due to the potential configuration only. Finally, the two potentials differ in their ranges. In contrast to the long-range Coulomb potential with an infinite energy spectrum, we find that the number of bound states for this short-range potential is finite.

The potential (10) has a richer structure than what we have noted above. In Appendix B, we find the conditions on the potential parameters such that the potential function crosses the real line at two points and thus will have two local extrema, a minimum and a maximum. Figure 3(a) shows such a structure in which the potential can support bound states as well as resonances. On the other hand, the potential configuration in Figure 3(b) may produce resonances but cannot support bound states.

## 3. TRA solution

To give the exact solution of the problem, we need to identify all ingredients in the wave function $|\psi(r)\rangle=\sum_{n=0}^{N}f_n|\phi_n(x)\rangle$. Since the basis elements $\{\phi_n(x)\}$ as given by Eq. (3) are now fully determined as shown in section 2 above, we need only to find an exact realization of the expansion coefficients of the wavefunction $\{f_n\}_{n=0}^{N}$. To do that, we start by substituting (8) into (6) with $U(x)=A(x-1)+(1-x^2)(B-Cx)$, $\mu^2-\nu^2=2A$ and $\mu^2+\nu^2=-2(\varepsilon+A)$. Subsequently, we use the three-term recursion relation of the Jacobi polynomial (A4) to obtain

$$J\phi_n(x) = \frac{\lambda^2}{2} c_n (1-x^2)(x-1)^{\frac{\mu}{2}}(x+1)^{\frac{\nu}{2}} \times$$

$$\left\{\left[\left(n+\frac{\mu+\nu+1}{2}\right)^2 - \frac{1}{4} + B - C\frac{\nu^2-\mu^2}{(2n+\mu+\nu)(2n+\mu+\nu+2)}\right] P_n^{(\mu,\nu)}(x)\right. \tag{12a}$$

$$\left. -2C\left[\frac{(n+\mu)(n+\nu)}{(2n+\mu+\nu)(2n+\mu+\nu+1)} P_{n-1}^{(\mu,\nu)} + \frac{(n+1)(n+\mu+\nu+1)}{(2n+\mu+\nu+1)(2n+\mu+\nu+2)} P_{n+1}^{(\mu,\nu)}\right]\right\}$$

This could be rewritten in terms of the basis function $\{\phi_n(x)\}$ as follows

$$J\phi_n(x) = \frac{\lambda^2}{2}(1-x^2)\left\{\left[\left(n+\frac{\mu+\nu+1}{2}\right)^2 - \frac{1}{4} + B - CF_n\right]\phi_n(x) - C[D_{n-1}\phi_{n-1}(x) + D_n\phi_{n+1}(x)]\right\}, \tag{12b}$$

where $F_n = \frac{\nu^2-\mu^2}{(2n+\mu+\nu)(2n+\mu+\nu+2)}$ and $D_n = \frac{2}{2n+\mu+\nu+2}\sqrt{\frac{(n+1)(n+\mu+1)(n+\nu+1)(n+\mu+\nu+1)}{(2n+\mu+\nu+1)(2n+\mu+\nu+3)}}$. Now, we insert the wavefunction series $|\psi\rangle = \sum_n f_n |\phi_n\rangle$ in the wave equation $J|\psi\rangle = 0$ and use (12b) to obtain the following symmetric three-term recursion relation for the expansion coefficients

$$\frac{B}{C}f_n = \left\{-C^{-1}\left[\left(n+\frac{\mu+\nu+1}{2}\right)^2 - \frac{1}{4}\right] + F_n\right\}f_n + D_{n-1}f_{n-1} + D_n f_{n+1}. \tag{13}$$

The potential parameters $B$ and $C$ are shown explicitly whereas the other parameter $A$ and the energy are implicit in $\mu$ and $\nu$ as shown by Eq. (11). We compare this recursion relation to that of the orthogonal polynomial $H_n^{(\mu,\nu)}(z^{-1};\theta,\sigma)$ shown as Eq. (9) in Ref. [13]. However, in that recursion relation, we make the change $\theta \to i\theta$, $z \to iz$ and write it in terms of the normalized version of the polynomial to obtain

$$(\cosh\theta) H_n^{(\mu,\nu)}(z^{-1};\theta,\sigma) = \left\{z^{-1}\sinh\theta\left[\left(n+\frac{\mu+\nu+1}{2}\right)^2 + \sigma\right] + F_n\right\} H_n^{(\mu,\nu)}(z^{-1};\theta,\sigma)$$
$$+ D_{n-1} H_{n-1}^{(\mu,\nu)}(z^{-1};\theta,\sigma) + D_n H_{n+1}^{(\mu,\nu)}(z^{-1};\theta,\sigma) \tag{14}$$

where $\theta \geq 0$. Comparing (13) and (14) gives $\cosh\theta = B/C$, $\sigma = -\frac{1}{4}$ and $z = -\sqrt{B^2 - C^2}$. These associations dictate that $B \geq C$, which in fact a necessary (but not sufficient) condition to produce a potential configuration that can support bound states as shown in Figure 2 and Figure 3(a). Therefore, we can write the expansion coefficients of the radial wavefunction (modulo an overall normalization constant) as

$$f_n(\varepsilon, A, B, C) = H_n^{(\mu,\nu)}\left(z^{-1};\theta,-1/4\right), \tag{15}$$

with $\mu$ and $\nu$ as given by Eq. (11). This completes the characterization of the exact solution of the problem. The wavefunction is given as the finite series $|\psi(r)\rangle = \sum_{n=0}^{N} f_n |\phi_n(x)\rangle$, where $N$ is the largest integer less than or equal to $\sqrt{-A/2} - \frac{1}{2}$. The energy eigenvalues are obtained from the spectrum formula of the orthogonal polynomial $H_n^{(\mu,\nu)}(z^{-1};\theta,-1/4)$ or the asymptotics of its continuous version as explained in Refs. [11,14,15]. Unfortunately, the analytic properties of these polynomials are not found in the mathematics literature. Up to date, it is still an open problem in orthogonal polynomials. This is in spite of the fact that these polynomials could be written explicitly to all degrees using the recursion relation (14) and the initial values $H_0^{(\mu,\nu)} = 1$, $H_{-1}^{(\mu,\nu)} \equiv 0$, albeit not in a closed from. For example, we obtain here

$$H_0^{(\mu,\nu)} = 1, \tag{16a}$$

$$H_1^{(\mu,\nu)} = \frac{B + G_0 - CF_0}{CD_0}, \tag{16b}$$

$$H_2^{(\mu,\nu)} = \frac{1}{CD_1}\left[(B + G_1 - CF_1)H_1^{(\mu,\nu)} - CD_0\right], \tag{16c}$$

$$H_3^{(\mu,\nu)} = \frac{1}{CD_2}\left[(B + G_2 - CF_2)H_2^{(\mu,\nu)} - CD_1 H_1^{(\mu,\nu)}\right], \tag{16d}$$

$$H_4^{(\mu,\nu)} = \frac{1}{CD_3}\left[(B + G_3 - CF_3)H_3^{(\mu,\nu)} - CD_2 H_2^{(\mu,\nu)}\right]. \tag{16e}$$

…. and so on,

where $G_n = \frac{1}{4}(2n+\mu+\nu)(2n+\mu+\nu+2)$. In the absence of knowledge of the analytic properties of the orthogonal polynomials $H_n^{(\mu,\nu)}(z^{-1};\theta,\sigma)$, we resort to numerical techniques to obtain the energy spectrum. Due to the energy dependence of the basis elements via the parameters $\mu$ and $\nu$, we may use one of two numerical procedure (i) the "*potential parameter spectrum* (PPS)" techniques or (ii) the direct diagonalization of the Hamiltonian matrix (HMD) in an energy independent basis similar to those in (3). In the latter, we use Gauss quadrature integral approximation associated with the Jacobi polynomials. We will not dwell on the details of these two techniques. Interested readers may consult Refs. [7,9,16] for a description of the PPS technique. In Appendix C, we a give a brief outline of how to obtain the matrix elements of the Hamiltonian and compute the energy spectrum using Gauss quadrature integral approximation [17,18]. Table 1 gives the full energy spectrum associated with the potential (10) for a given set of values of the parameters A, B and C. The Table shows the rate of convergence of the energy spectrum with an increase in the basis size. In Figure 4, we plot the bound state wave functions associated with the energy spectrum in Table 1. The wavefunction of a bound state with energy $E_k$, is calculated by the finite series

$$\psi_k(r) \sim (\coth \lambda r - 1)^{\frac{1}{2}\mu_k}(\coth \lambda r + 1)^{\frac{1}{2}\nu_k} \sum_{n=0}^{k} c_n^k f_n(\varepsilon_k, A, B, C) P_n^{(\mu_k,\nu_k)}(\coth \lambda r), \tag{17}$$

where $\varepsilon_k = 2E_k/\lambda^2$, $\mu_k = \sqrt{-\varepsilon_k}$, $\nu_k = -\sqrt{-\varepsilon_k - 2A}$ and

$$c_n^k = \sqrt{(2n+\mu_k+\nu_k+1)\frac{\Gamma(n+1)\Gamma(n+\mu_k+\nu_k+1)}{\Gamma(n+\mu_k+1)\Gamma(n+\nu_k+1)}}. \tag{18}$$

## 5. Conclusion and discussion

In this work, we used the TRA to obtain the bound states solution (energy spectrum and wavefunctions) of the S-wave Schrödinger equation for a generalization of the hyperbolic Eckart potential with and added short-range $1/r^3$ singular potential. To the best of our knowledge, this is the first time that a solution of such singular potential is obtained analytically. The solution is a finite series of square integrable functions with orthogonal polynomials as expansion coefficients of the series. The analytic properties of these polynomials are yet to be derived and we call upon experts in the field of orthogonal polynomials to address this open problem [19]. Currently, we are using the results obtained in this work to study the binding of an electron to a molecule with an effective electric quadrupole moment. This was made possible because the electric quadrupole potential goes radially as $1/r^3$. Nonetheless, the short-range behavior of the potential (10) in such a study could be attributed physically to the screening of the electric interaction by the molecular electron cloud.

## Acknowledgements

This work is supported by the Saudi Center for Theoretical Physics (SCTP). We are grateful to Prof. H. Bahlouli (of KFUPM) for proof reading the original version.

## Appendix A: The Jacobi polynomial

For ease of reference, we list below the basic properties of the Jacobi polynomial that could be found in many places including Ref. [20]

$$P_n^{(\mu,\nu)}(x) = \frac{\Gamma(n+\mu+1)}{\Gamma(n+1)\Gamma(\mu+1)} {}_2F_1(-n, n+\mu+\nu+1; \mu+1; \tfrac{1-x}{2}) = (-)^n P_n^{(\nu,\mu)}(-x). \quad (A1)$$

where $\mu > -1$ and $\nu > -1$ for $x \in [-1, +1]$. It satisfies the following differential equation

$$\left\{ (1-x^2)\frac{d^2}{dx^2} - \left[(\mu+\nu+2)x + \mu-\nu\right]\frac{d}{dx} + n(n+\mu+\nu+1) \right\} P_n^{(\mu,\nu)}(x) = 0, \quad (A2)$$

and the following differential relation

$$(1-x^2)\frac{d}{dx} P_n^{(\mu,\nu)} = -n\left(x + \frac{\nu-\mu}{2n+\mu+\nu}\right) P_n^{(\mu,\nu)} + 2\frac{(n+\mu)(n+\nu)}{2n+\mu+\nu} P_{n-1}^{(\mu,\nu)}. \quad (A3)$$

It also satisfies the following three-term recursion relation

$$\begin{aligned} x P_n^{(\mu,\nu)} &= \frac{\nu^2-\mu^2}{(2n+\mu+\nu)(2n+\mu+\nu+2)} P_n^{(\mu,\nu)} \\ &+ \frac{2(n+\mu)(n+\nu)}{(2n+\mu+\nu)(2n+\mu+\nu+1)} P_{n-1}^{(\mu,\nu)} + \frac{2(n+1)(n+\mu+\nu+1)}{(2n+\mu+\nu+1)(2n+\mu+\nu+2)} P_{n+1}^{(\mu,\nu)} \end{aligned} \quad (A4)$$

The associated orthogonality relation reads as follows

$$\int_{-1}^{+1}(1-x)^{\mu}(1+x)^{\nu}P_n^{(\mu,\nu)}(x)P_m^{(\mu,\nu)}(x)dx = \frac{2^{\mu+\nu+1}}{2n+\mu+\nu+1}\frac{\Gamma(n+\mu+1)\Gamma(n+\nu+1)}{\Gamma(n+1)\Gamma(n+\mu+\nu+1)}\delta_{nm} \quad (A5)$$

On the other hand, in Ref. [21] the Authors derive the following orthogonality for $x \geq +1$ and with $\mu+\nu < -2N-1$ and $\mu > -1$

$$\int_{1}^{\infty}(x-1)^{\mu}(x+1)^{\nu}P_n^{(\mu,\nu)}(x)P_m^{(\mu,\nu)}(x)dx = \frac{2^{\mu+\nu+1}}{2n+\mu+\nu+1}\frac{\Gamma(n+\mu+1)\Gamma(n+\nu+1)}{\Gamma(n+1)\Gamma(n+\mu+\nu+1)}\frac{\sin\pi\nu}{\sin\pi(\mu+\nu+1)}\delta_{nm}. \quad (A6)$$

and $n,m \in \{0,1,2,...,N\}$. Equivalently,

$$\int_{1}^{\infty}(x-1)^{\mu}(x+1)^{\nu}P_n^{(\mu,\nu)}(x)P_m^{(\mu,\nu)}(x)dx = \frac{(-1)^{n+1}2^{\mu+\nu+1}}{2n+\mu+\nu+1}\frac{\Gamma(n+\mu+1)\Gamma(n+\nu+1)\Gamma(-n-\mu-\nu)}{\Gamma(n+1)\Gamma(-\nu)\Gamma(\nu+1)}\delta_{nm}. \quad (A6)'$$

Moreover, if we make the replacement $x \to 2x+1$ in (A6), we obtain

$$\int_{0}^{\infty}x^{\mu}(x+1)^{\nu}P_n^{(\mu,\nu)}(2x+1)P_m^{(\mu,\nu)}(2x+1)dx = \frac{1}{2n+\mu+\nu+1}\frac{\Gamma(n+\mu+1)\Gamma(n+\nu+1)}{\Gamma(n+1)\Gamma(n+\mu+\nu+1)}\frac{\sin\pi\nu}{\sin\pi(\mu+\nu+1)}\delta_{nm}, \quad (A7)$$

where $\mu+\nu < -2N-1$, $\mu > -1$ and $n,m \in \{0,1,2,...,N\}$.

## Appendix B: The rich potential structure

In this Appendix, we show that under the right conditions the potential (10) displays a richer structure by having two local extrema (minimum and maximum). To find these conditions, it is easier to deal with $U(x)$ rather than $V(r)$. This is possible because there is one-to-one correspondence between $r$ and $x$. Thus, the potential vanishes if

$$(x-1)[A+(x+1)(Cx-B)] = 0, \quad (B1)$$

which is satisfies at $x = 1$ (i.e., $r \to \infty$) and at

$$x_{\pm} = \frac{1}{2}\left[\gamma - 1 \pm \sqrt{(\gamma+1)^2 - 4\xi}\right], \quad (B2)$$

where $\gamma = B/C$ and $\xi = A/C$. Thus, the potential crosses the real line at $r_{\pm} = \lambda^{-1}\coth^{-1}(x_{\pm})$ if and only if $4\xi < (\gamma+1)^2$ and $x_{\pm} \geq +1$. Very often, the first of these two conditions is satisfied but not the second. In that case, the potential will have only one extremum as shown in Figure 2 or none. These local extrema correspond to real solutions of the equation $dU/dx = 0$, which gives $3x^2 - 2\gamma x + \xi - 1 = 0$ whose solution is

$$\tilde{x}_{\pm} = \frac{1}{3}\left[\gamma \pm \sqrt{\gamma^2 + 3(1-\xi)}\right]. \tag{B3}$$

In such cases, the two potential extrema are located at $\tilde{r}_{\pm} = \lambda^{-1}\coth^{-1}(\tilde{x}_{\pm})$ and their values are $V(\tilde{r}_{\pm})$. Figure 3(a) shows a configuration corresponding to $\gamma = 7$ and $\xi = 17$. In this case, one would expect the existence of resonances in addition to bound states. However, in other cases, only resonances occur but no bound states. This corresponds to the existence of two extrema but without potential crossing as shown in Figure 3(b) for $\gamma = 7$ and $\xi = 16.2$.

## Appendix C: Hamiltonian matrix, Gauss quadrature and evaluation of the energy spectrum

The action of the Hamiltonian operator on the basis element (3) is obtained from Eq. (6) by setting $\varepsilon = 0$ and substituting the potential (10) that reads $U(x) = A(x-1) + (1-x^2)(B-Cx)$, giving

$$H\phi_n(x) = -\frac{\lambda^2}{2}c_n(1-x^2)(x-1)^\alpha (x+1)^{-\beta} \times$$

$$\left\{\left(\frac{\mu-2\alpha}{1-x} - \frac{\nu+2\beta}{1+x}\right)\left[2\frac{(n+\mu)(n+\nu)}{2n+\mu+\nu}P_{n-1}^{(\mu,\nu)} - n\left(x + \frac{\nu-\mu}{2n+\mu+\nu}\right)P_n^{(\mu,\nu)}\right]\right. \tag{C1}$$

$$\left. +\left[\frac{1}{2}\frac{4\alpha^2}{1-x} + \frac{1}{2}\frac{4\beta^2 + A}{1+x} - (\alpha-\beta)(\alpha-\beta+1) - n(n+\mu+\nu+1) - B + Cx\right]P_n^{(\mu,\nu)}\right\}$$

Choosing $2\alpha = \mu$, $2\beta = -\nu$ and using the integral transform $\lambda \int_0^\infty \ldots dr = -\int_1^\infty \ldots \frac{dx}{1-x^2}$, we obtain the following matrix elements of the Hamiltonian

$$\langle \phi_m | H | \phi_n \rangle = \lambda \int_0^\infty \phi_m(x) H \phi_n(x) dr$$
$$= \frac{\lambda^2}{2}c_m c_n \int_1^\infty (x-1)^\mu (x+1)^\nu W(x) P_m^{(\mu,\nu)}(x) P_n^{(\mu,\nu)}(x) dx \tag{C2}$$

where $W(x) = \frac{1}{2}\frac{\mu^2}{1-x} + \frac{1}{2}\frac{\nu^2 + A}{1+x} - \left(n + \frac{\mu+\nu+1}{2}\right)^2 + \frac{1}{4} - B + Cx$. The recursion relation and orthogonality of the Jacobi polynomial show that all terms in $W(x)$ that are proportional to $x^k$, where $k$ is a non-negative integer, will produce exact matrix elements that are $2k+1$ diagonal (i.e., a banded diagonal matrix whose diagonal band is of width $2k+1$). Therefore, only the first two terms in $W(x)$ will produce a matrix with non-zero entries everywhere, which needs to be evaluated numerically. Note that we can eliminate one or both of these terms by choosing $\mu^2 = 0$ and/or $\nu^2 = -A$. However, as will be shown below this will severely limit our calculation and may lead to inconsistencies. Therefore, we keep both parameters arbitrary for the moment.

To calculate the matrix elements that represent the first two terms in $W(x)$, we use Gauss quadrature integral approximation associated with the Jacobi polynomial. The details of this method are found in Refs. [17,18]. The essence of the method might be summarized as follows. Let $X_{n,m}$ be the matrix elements obtained by choosing $W(x) = 2x/\lambda^2$ in Eq. (C2). Then by using the recursion relation and orthogonality of the Jacobi polynomial it is easy to show that $X$ is the following tridiagonal symmetric matrix

$$X_{n,m} = F_n \delta_{n,m} + D_n \delta_{n,m-1} + D_{n-1} \delta_{n,m+1}, \qquad (C3)$$

where the numbers $\{F_n, D_n\}$ are defined in section 3 below Eq. (12b). Now, let $\{\tau_n\}_{n=0}^{N}$ be the eigenvalues of the $(N+1) \times (N+1)$ truncated version of the matrix $X$ and $\{\Lambda_{m,n}\}_{m=0}^{N}$ be its normalized eigenvector associated with the eigenvalue $\tau_n$. Then an approximate evaluation of the matrix elements that represent any function $W(x)$ is

$$W_{n,m} \simeq \frac{\lambda^2}{2} \left( \Lambda \cdot \Omega \cdot \Lambda^{\text{T}} \right)_{n,m}, \qquad (C4)$$

where $\Omega$ is a diagonal matrix with diagonal elements $\Omega_{n,n} = W(\tau_n)$. Therefore, the evaluated Hamiltonian matrix has finally the following elements

$$\frac{2}{\lambda^2} H_{n,m} \simeq \left[ \frac{1}{4} - B - \left( n + \frac{\mu+\nu+1}{2} \right)^2 \right] \delta_{n,m} + C X_{n,m}$$
$$+ \frac{\mu^2}{2} \left( \Lambda \cdot \Omega_- \cdot \Lambda^{\text{T}} \right)_{n,m} + \frac{\nu^2 + A}{2} \left( \Lambda \cdot \Omega_+ \cdot \Lambda^{\text{T}} \right)_{n,m} \qquad (C5)$$

where $(\Omega_\pm)_{n,m} = \frac{\delta_{n,m}}{1 \pm \tau_n}$. The size of this matrix (number of basis elements) is $N+1$ with $n, m \in \{0, 1, 2, ..., N\}$, $\mu > -1$ and $\mu + \nu < -2N - 1$. To calculate the energy spectrum, we write the energy eigenvalue equation (wave equation) $H|\psi\rangle = E|\psi\rangle$ in the energy independent basis as $\sum_m f_m H |\phi_m\rangle = E \sum_m f_m |\phi_m\rangle$. Projecting from left by $\langle \phi_n |$, we obtain $\sum_m H_{n,m} f_m = E \sum_m \omega_{n,m} f_m$, where $\omega$ is the basis overlap matrix whose elements are $\omega_{n,m} = \langle \phi_n | \phi_m \rangle$. Consequently, the energy spectrum is calculated as the generalized eigenvalues of the matrix equation $H|f\rangle = E(\omega|f\rangle)$. To obtain the overlap matrix $\omega$, we follow the same Gauss quadrature procedure outlined above giving

$$\omega_{n,m} \simeq -\left( \Lambda \cdot \tilde{\Omega} \cdot \Lambda^{\text{T}} \right)_{n,m}, \qquad (C6)$$

where $\tilde{\Omega}_{n,m} = \frac{\delta_{n,m}}{1 - \tau_n^2}$.

Finally, we may attempt to ease the calculation process and speed it up by eliminate one of the two unphysical computational parameters $\mu$ or $\nu$. A good way to do that might be to use the energy independent relation used below Eq. (9): $\mu^2 - \nu^2 = 2A$. To decide on which one to eliminate, we exploit the fact that at the right value of these parameters their relation to the energy is $\mu^2 + \nu^2 = -2(\varepsilon + A)$. Thus, $\frac{\partial \varepsilon}{\partial \mu} = -\mu < 1$ and $\frac{\partial \varepsilon}{\partial \nu} = -\nu > 2N$. This implies that if we increase the basis size to improve accuracy then the rate of change of the spectrum with respect to variations in the parameter $\nu$ becomes large. Therefore, we eliminate $\nu$ in favor of $\mu$ by setting $\nu = -\sqrt{\mu^2 - 2A}$ for a give value of $\mu$. However, the condition $\mu + \nu < -2N - 1$ dictates that $N < \frac{1}{2}\left(\sqrt{\mu^2 - 2A} - \mu - 1\right)$, which restricts severely the size of the matrices and hinder the accuracy of the calculation. In conclusion, we keep both parameters arbitrary and follow a strategy whereby we start with a given basis size, $N$, then choose as small values of $\mu$ and $\nu$ as possible without violating the constraints $\mu > -1$ and $\mu + \nu < -2N - 1$. To finish, we vary the value of $\mu$ while keeping $\nu$ fixed until a plateau of computational stability of the energy spectrum is reached for large enough $N$. Figure 5 shows the variation in the computed values of the energy spectrum Of Table 1 as we vary $\mu$ while keeping $\nu$ fixed for $N = 100$. The plateau of computational stability (range of values of $\mu$ with no significant change in the result) is evidently larger for lower bound states.

# Table Caption

**Table 1**: The complete finite bound states energy spectrum for the potential (10) in units of $-\frac{1}{2}\lambda^2$. The potential parameters are taken as $A = -300$, $B = 5$ and $C = 3$. We show the rate of convergence of the calculation with an increase in the basis size $N$. The computational parameters $\mu$ and $\nu$ are chosen after making the *plateau of stability* analysis as explained at the end of Appendix C. These are $\mu = 1.5$ and $\nu = -2N - \mu - 2.0$.

### Table 1

| $n$ | $N = 10$ | $N = 20$ | $N = 50$ | $N = 100$ |
|---|---|---|---|---|
| 0 | 249.6186960 | 249.6474349 | 249.6474353 | 249.6474353 |
| 1 | 121.1023091 | 121.1387777 | 121.1387781 | 121.1387781 |
| 2 | 54.5612094 | 54.5922339 | 54.5922342 | 54.5922342 |
| 3 | 20.1791388 | 20.1738603 | 20.1738321 | 20.1738321 |
| 4 | 4.8218491 | 4.2733151 | 4.2434960 | 4.2427578 |

# Figures Captions

**Fig. 1**: The function $\mu(\varepsilon) + \nu(\varepsilon)$ (solid curve) as given by Eq. (11) for $A = -50$. The horizontal dashed lines correspond to $-2k - 1$ for $k = 0, 1, 2, ..$ whereas the horizontal solid line is $-2N - 1$. The condition $\mu + \nu < -2N - 1$ gives the maximum value of $N$ and shows the allowed ranges of energy.

**Fig. 2**: Plots of the potential function (10) in units of $\frac{1}{2}\lambda^2 C$ for the ratio $B/C = 2.0$ and for several values of the ratio $A/C = \{-2.0, 0.0, 2.0, 4.0\}$ from top to bottom. Bound states require $B \geq C$.

**Fig. 3**: Plots of the potential function (10) in units of $\frac{1}{2}\lambda^2 C$ for the ratio $B/C = 7.0$ and with $A/C = 17.0$ in part (a) and $A/C = 16.2$ in part (b).

**Fig. 4**: Plot of the un-normalized bound state wavefunctions associated with the energy spectrum in Table 1. These were calculated using Eq. (17).

**Fig. 5**: Changes in the calculation of the bound states energies of Table 1 (for a basis size of $N = 100$) as $\mu$ varies while keeping $\nu$ fixed at $\nu = -2N - \mu - 2.0$. The horizontal axis is $\mu$ and $\Delta$ is an average measure of the energy variations (in units of $\frac{1}{2}\lambda^2$) within the plateau, which gives a measure of accuracy. The Figure shows that a good choice of value of $\mu$ for the whole spectrum is within the range 1.0 to 2.0.

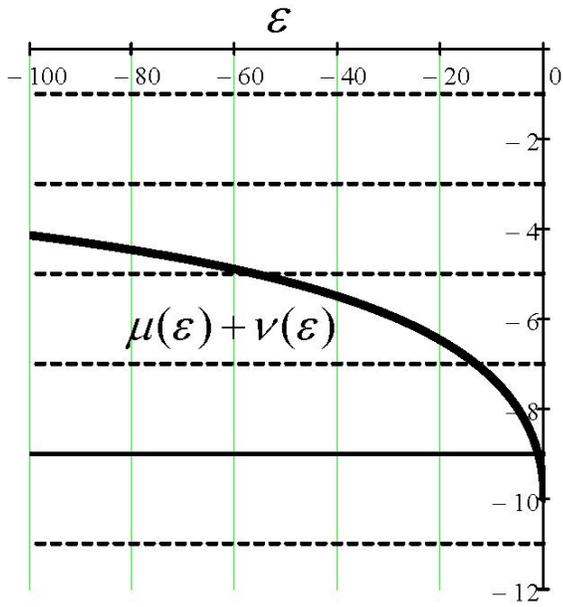 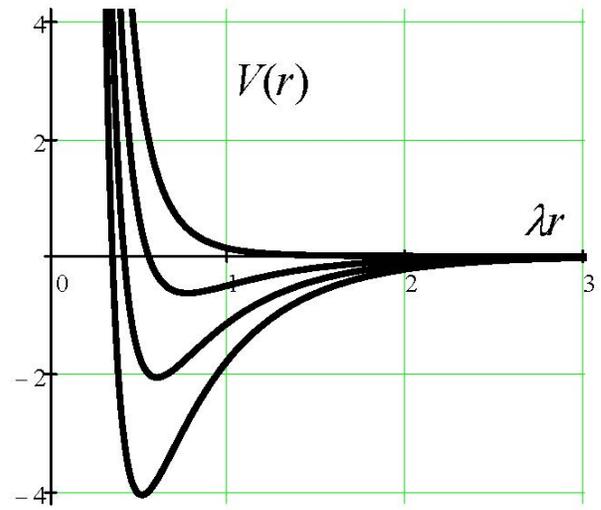

**Fig. 1**  **Fig. 2**

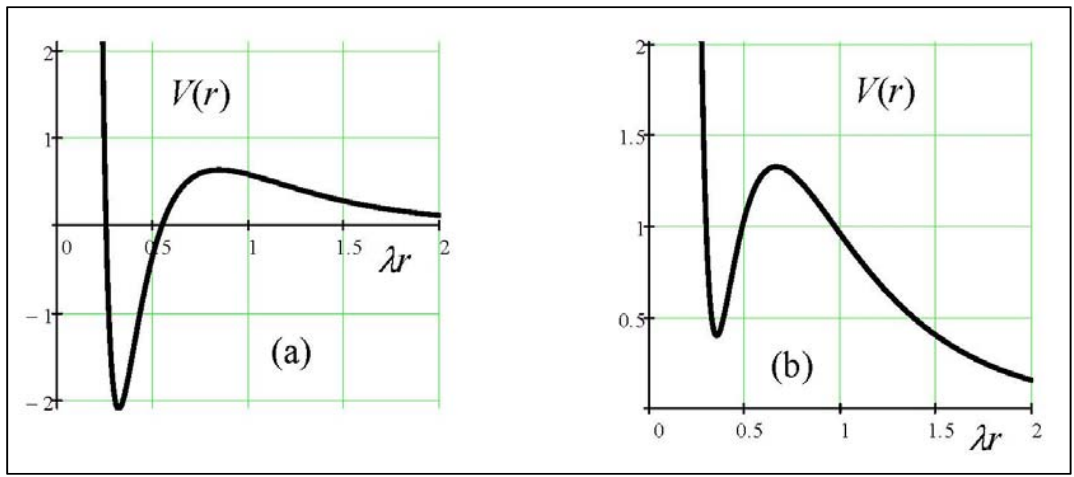

**Fig. 3**

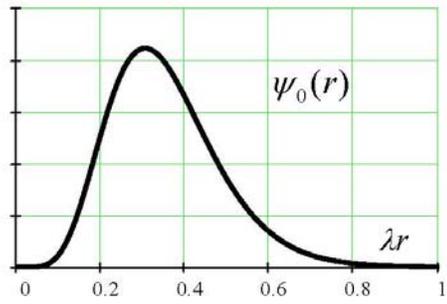
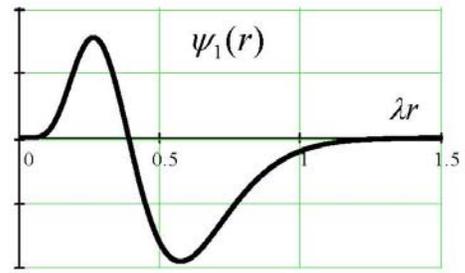
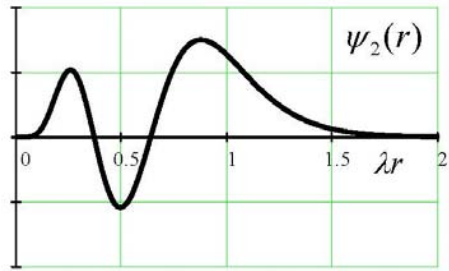
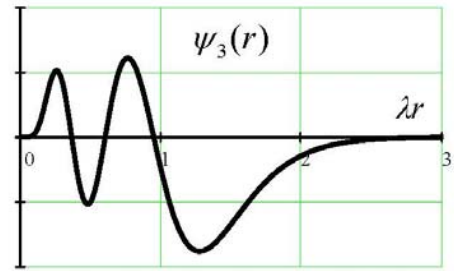
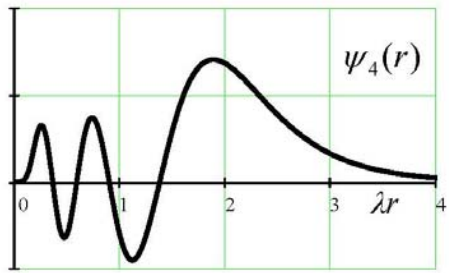

**Fig. 4**

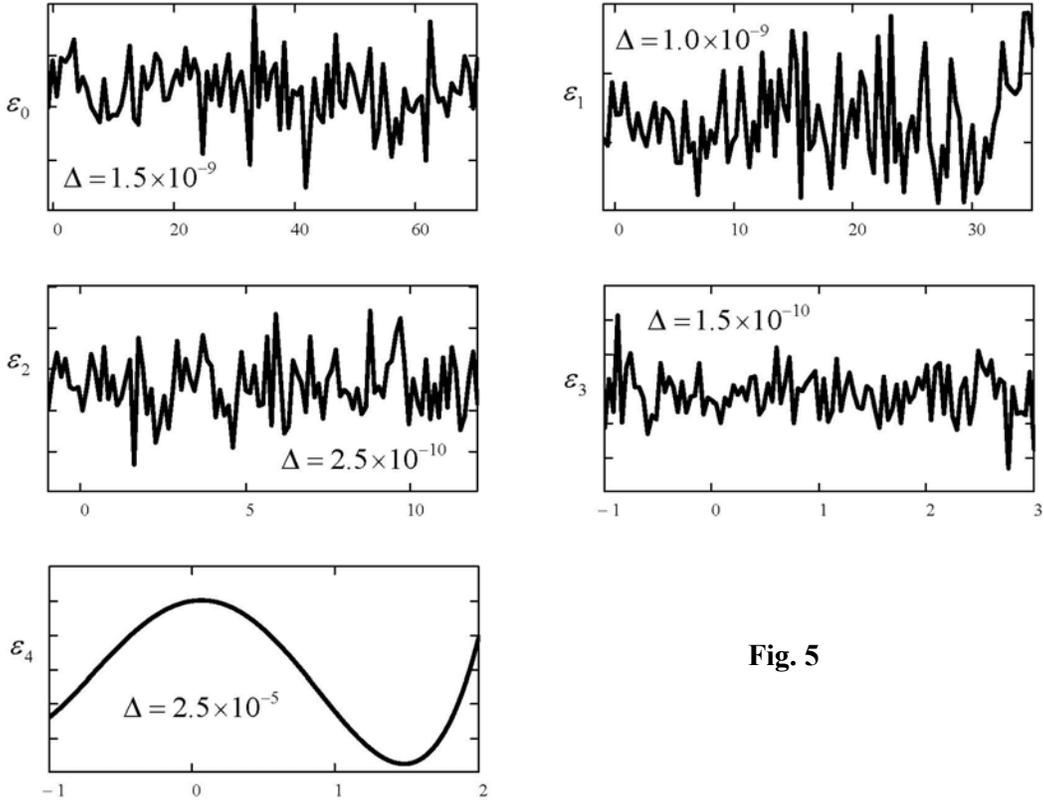

**Fig. 5**